\documentclass[journal]{IEEEtran}
\usepackage{amssymb,amsmath,latexsym,amscd,amsfonts,latexsym,mathtools,mathrsfs,array,bm}
\usepackage[usenames,dvipsnames]{xcolor}
\usepackage[colorlinks,linkcolor=Black,citecolor=Black,urlcolor=Black]{hyperref}
\usepackage{cite}
\usepackage[most]{tcolorbox}
\usepackage{empheq}

\usepackage{graphicx}
\usepackage[caption=false,font=footnotesize]{subfig}
\usepackage{setspace}
\usepackage[bottom ,hang, flushmargin]{footmisc}
\usepackage{soul}

\usepackage{dblfloatfix, url}

\newcommand{\ud}{\,\mathrm{d}}
\newcommand{\ket}[1]{|#1\rangle}
\newcommand{\bra}[1]{\langle#1|}

\begin{document}
%\title{Efficient Analysis of Phoxonic Crystal Slabs}
\title{Efficient Analysis of Confined Guided Modes in Phoxonic Crystal Slabs}

\author{Mohammad~Hasan~Aram
        and ~Sina~Khorasani%,~\IEEEmembership{Senior Member,~IEEE}
        % <-this % stops a space
\thanks{M. H. Aram is with the School
of Electrical Engineering, Sharif University of Technology, Tehran,
Iran (e-mail: mharam@ee.sharif.ir).}% <-this % stops a space
\thanks{S. Khorasani is with the \'Ecole Polytechnique F\'ed\'erale de Lausanne (EPFL),
CH-1015 Lausanne, Switzerland, on leave from School of Electrical Engineering, Sharif University of Technology, Tehran, Iran (e-mail: sina.khorasani@epfl.ch).}% <-this % stops a space
\thanks{This work has been supported in part by Iranian National Science Foundation (INSF) under the grant 93026841. M.H.A. is being supported through a postdoctoral fellowship at Sharif University of Technology. S.K. is being supported by the Research Deputy of Sharif University of Technology over a sabbatical visit, as well as Laboratory of Photonics and Quantum Measurements (LPQM) at EPFL.}
%\thanks{Manuscript received March 17, 2017, revised June 9, 2017.}
}

\maketitle

\begin{abstract}
Today's standard fabrication processes are just capable of manufacturing slab of photonic and phononic crystals, so an efficient method for analysis of these crystals is indispensable. Plane wave expansion (PWE) as a widely used method in studying photonic and phononic (phoxonic) crystals in full three dimensions is not suitable for slab analysis in its standard form, because of convergence and stability issues. Here, we propose a modification to this method which overcomes these limitations. This improved method can be utilized for calculation of both photonic and phononic modes in phoxonic slabs. While in the standard three-dimensional PWE, Fourier series are used to estimate the field dependence across the normal component of the slab, we expand the fields across the third dimension using eigenmodes of a plain unstructured slab. Despite its approximate nature, this approach is observed to be both much faster and more accurate than the conventional PWE method and can give a very accurate estimation of confined propagating modes. As an application example of the proposed method we investigate a non-reciprocal photonic device. This device is a phoxonic slab waveguide which changes modes of optical waves by elastic waves and can be used as an optical insulator or mode converter.
\end{abstract}

\begin{IEEEkeywords}
Phononic crystal, Photonic crystal, Phoxonic crystal, Plane Wave Expansion, Photo-elastic interaction.
\end{IEEEkeywords}

% \fi
%\IEEEpeerreviewmaketitle

\section{Introduction}

\IEEEPARstart{T}{here} exist many different types of waves in the nature, among them elastic and electromagnetic waves being the most famous and known ones. Because of the availability of sound waves as the scalar polarization of elastic pressure waves, their behavior was well known to physicists many years before the formulation of Maxwell's equations. With the advent of electromagnetism and special relativity, the focus of study in wave phenomena shifted primarily to the realm of optics and electromagnetics and consequently photonic crystals (PtC) were introduced \cite{Yablonovitch, John1} before their phononic counterparts \cite{Narayanamurti}. They are called crystals since their constituents are periodic. Periodicity of permittivity in photonic crystals, and density and stiffness in phononic crystals (PnC) may be held responsible for their characteristics. The main property of these crystals is their ability to prevent some continuous frequency ranges of waves from propagating inside them. These frequency ranges are called forbidden photonic/phononic frequency gaps, which are caused by destructive interference of waves inside the crystal.

Because of this interesting property, photonic and phononic crystals have found many applications in different fields of science and technology \cite{Olsson}, among them are optomechanics \cite{Taofiq,Hill,Aspelmeyer,Aspelmeyer2,KALAEE}, sensors \cite{Oseev}, acoustic and optical insulators, waveguides, filters and high $Q$ cavities \cite{Khelif,Djafari,Safavi,Khelif2,Torres,Mohammadi}. With the help of simultaneous phononic and photonic band gaps, one can create complete optical insulation \cite{Zongfu}. Acoustic insulators can be used to isolate gyroscopes resonators and oscillators from ambient noise and vibrations\cite{Soliman}. By utilizing acoustic waveguides, one can miniaturize acoustic delay elements which are used in signal processing and oscillators based on delay line. These waveguides can also be used in impedance matching and focusing in acoustic photography like medical ultrasound \cite{Olsson}.

By utilizing photonic crystals, power efficiency of solar cells \cite{Zhou,Mutitu,Zeng,Bermel} and extraction efficiency of light emitting diodes \cite{Ichikawa,Boroditsky,Fan} can be significantly improved. It is possible to embed a layer of PtC beneath miniaturized antennas manufactured on chips to increase the power of radiated waves through reflecting the waves, which would otherwise penetrate into the substrate \cite{Brown,Temelkuran}. PtC cavities can be used to construct ultra-fast low-threshold tiny lasers \cite{Hirayama,Englund}, which find use in the field of cavity quantum electrodynamics \cite{AramJMP} as well. Like acoustic waveguides, optical waveguides created in photonic crystals are also widely used \cite{Chutinan,Lin,Villeneuve,Loncar}.    

Numerous methods have been developed for PnCs analysis, most of them being in common with those of PtCs. The most well-known methods are plane wave expansion (PWE), finite element (FE) and finite difference time domain (FDTD) \cite{Reinke}. Among these methods, PWE is more popular than the other two since it is much simpler to set up and suitable for periodic structures. It is based on Bloch theorem and approximating periodic quantities by their Fourier series. Its shortcoming arises when the structure is not periodic nor invariant in one or more dimensions or the wave does not propagate uniformly throughout the whole crystal. For instance, PnC slab in not periodic nor invariant along its thickness or surface acoustic waves (SAWs) propagate near the surface of a semi-infinite crystal. In order to utilize this method in these circumstances we have to make some changes either in its standard formulation or in our structure. 

A reliable and conventional approach for dealing with PnC slabs is to sandwich the slab between two vacuum layers and create a 3D structure by stacking such sandwiches on top of each other\cite{Hou1}. The obtained structure is then analyzed with full 3D PWE method. Despite its accuracy, this method is time consuming and its convergence is normally non-uniform. Nevertheless, its accuracy is not desirable in solid-solid crystals of large elastic mismatch. To resolve this problem a new formulation has been developed based on inverse rule in contrast to conventional Laurent's rule to obtain Fourier coefficients \cite{Cao,Ghasemi}.

Sigalas and Economoua tried to combine PWE with plate theory to calculate PnC slabs band structure \cite{Sigalas1}. Although it is a good approach for thin slabs, its error increases as the slab becomes thicker. Tanaka and Tamura proposed a modified PWE method for SAWs analysis \cite{Tanaka}. They supposed surface waves to decay exponentially by getting away from the crystal surface. This idea was then utilized to study PnC slabs \cite{Charles} since they are similar to semi-infinite crystals as they also have free surfaces. It is not an efficient method for crystal slab analysis because it needs a numerical root searching procedure which is time consuming and not straightforward at all. 

There exists another variation of the PWE method for PnC slab analysis, proposed by Hsu and Wu, which is based on Mindlin's plate theory \cite{Hsu}. It has been claimed that this method is much faster than full 3D PWE method but it is restricted to thin slabs and low frequencies. Much research and effort has been put into this domain to overcome the limitations of PWE while avoiding coding complexity of fintie element and boundary element formalisms \cite{Nekuee,Shi,Qiu,Shi2}.

In this article we introduce a new modified PWE method which can be employed to find modes of both photonic and phononic modes of phoxonic crystal slabs which are composed of isotropic materials. It is shown to be much faster than the standard full 3D PWE method. Then we investigate optomechanical interactions in an optomechanical slab waveguide that is used to design a recently proposed non-reciprocal device \cite{Khorasani} and can be as well utilized as an optical insulator.

%%%%%%%%%%%%%%%%%%%%%%%%%%%%%
\section{Modified PWE Method}

PWE method is based on approximating wave profile by Fourier series. Although this approximation is inevitable for a real periodic crystal, it seems not to be so efficient for a crystal slab along its normal axis. In this section, we introduce functions which can replace Fourier series along the off-plane, or normal component to the slab. We furthermore suppose that the wave profile along the slab normal axis does not change significantly when we carve out some holes or insert some inclusions inside the slab, thus enabling the estimation of wave profile along the normal axis by that of a plain slab. Then, Fourier series is used to expand the waves along the two other orthogonal in-plane axes.  

\subsection{Phononic Crystal Slab}
\begin{figure}[ht]
\centering
\includegraphics[width=3.5in] {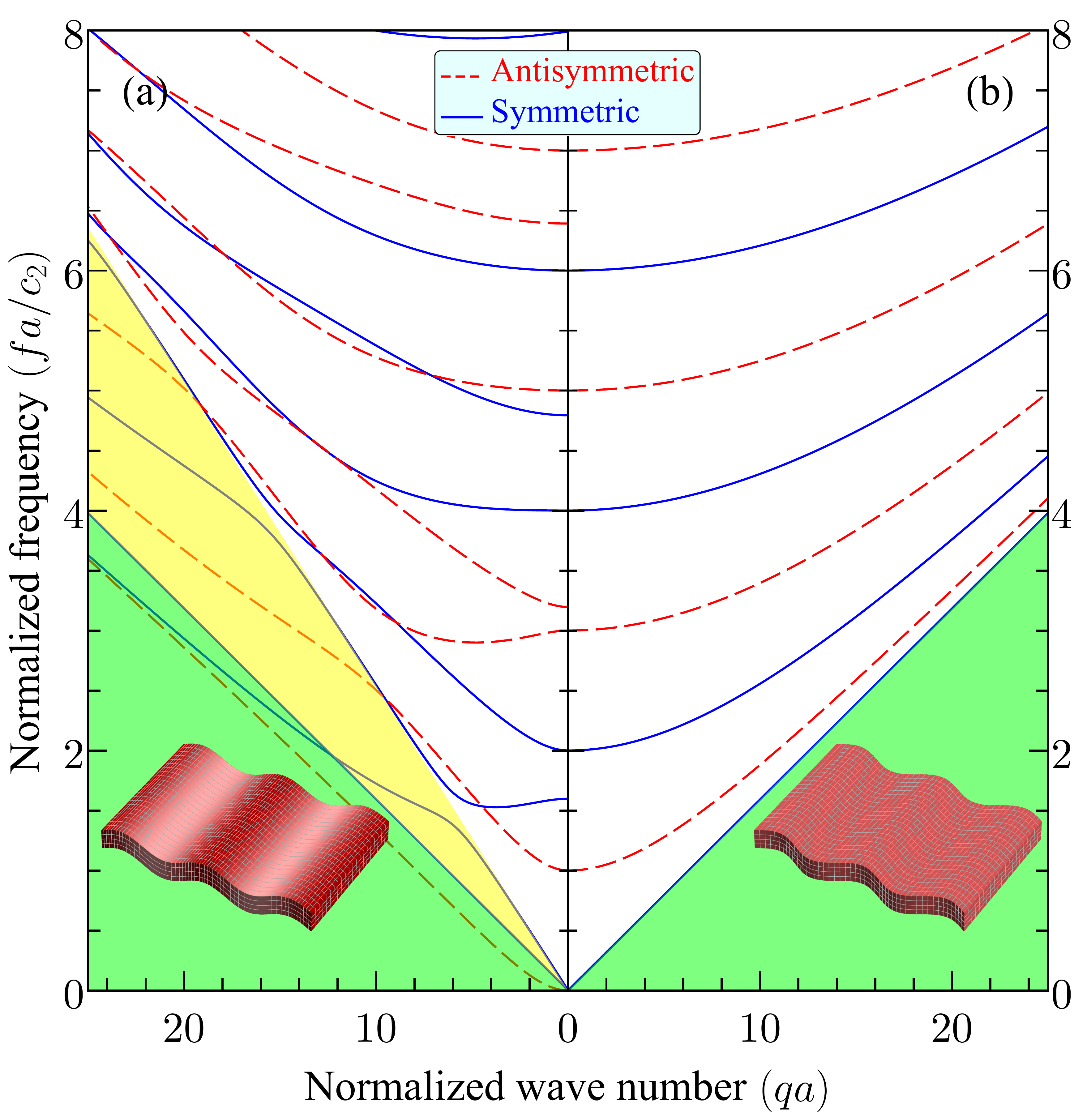}
\caption{Dispersion curves of an isotropic Silicon plate for (a) longitudinal coupled with vertical transverse and (b) horizontal transverse waves. Si Lam$\acute{\text{e}}$ constants and mass density equal $\lambda=44.27G\text{Pa}$, $\mu=80G\text{Pa}$ and $\rho=2329 k\text{g}/\text{m}^{3}$. Plate thickness is assumed to be $\text{t} = 0.5a$ where $a$ is an arbitrary length. Insets of the figure show both types of the waves propagating in the plate.}
\label{F1} 
\end{figure}

To begin, we first have to find modes of a plain elastic plate. As stated above, the plate is assumed to be made of isotropic material.
%Without loss of generality and for simplicity we assume the plate to be isotropic.
 In this case wave equation is written as \cite{Graff}
\begin{equation*} 
(\lambda + \mu) \nabla \nabla \cdot \mathbf{u} + \mu \nabla^2 \mathbf{u} = \rho \ddot{\mathbf{u}} ,
\end{equation*}  
which after expanding is equivalent to 
\begin{equation} \label{E1}
\sum_{j=1}^3 \left[(\lambda+\mu) \frac{\partial^2 u_j}{\partial x_j \partial x_i} + \mu \frac{\partial^2 u_i}{\partial x_j^2} \right] = \rho \ddot{u_i}, \quad (i=1,2,3).
\end{equation}  
In this equation $x_i$ and $u_i$ are the $i$'th component of position and displacement respectively,  $\lambda$ and $\mu$ are Lam$\acute{\text{e}}$ constants and $\rho$ is the mass density of the plate. It is simple to show that two kinds of waves can propagate inside this plate, longitudinal coupled with vertical transverse, and horizontal transverse. We assume the plate normal direction to be along the $z-$axis and the plate to be surrounded between $z=-\text{t}/2$ and $z=\text{t}/2$ surfaces. We also assume the wave to propagate along the $x-$axis. So we can write $ \mathbf{u}(\mathbf{r}) = \mathbf{u}(x,z) $. 
In the horizontal transverse wave, displacement is just along the $y-$axis and in the other one it is along $x-$ and $z-$axes. By solving wave equation, horizontal transverse wave is obtained as \cite{Graff}
\begin{equation} \label{E2}
u_y(x,z) = \left\{
\begin{array}{lr}
A_1 \cos (k z ) e^{-\imath(q x - \Omega t)} & \text{Symmetric}\\
A_2 \sin (k z) e^{-\imath(q x - \Omega t)} & \text{Anti-symmetric}
\end{array} \right.,
\end{equation}  
in which $\Omega = \sqrt{(k^2+ q^2) c_2^2}$ is the angular frequency, $c_2=\sqrt{\mu/\rho}$ is the phase velocity of transverse wave and $q$ is the wave number. Applying boundary conditions, we conclude,
\begin{equation*} 
k= n \pi /\text{t}, \quad n =0,1,2, \hdots . 
\end{equation*} 
The wave equation and boundary conditions give \cite{Graff}
\begin{subequations} \label{E3}
\begin{align} 
& u_x =  \left[A_2 \imath q \cos (k_1 z )  - B_1 k_2 \cos (k_2 z) \right] e ^{-\imath(q x - \Omega t)}, \label{E3a}  \\
& u_z = \left[ - A_2 k_1  \sin (k_1 z )  + B_1 \imath q \sin (k_2 z) \right] e ^{-\imath(q x - \Omega t)}, \label{E3b} \\
& \frac{\tan(k_2 \text{t}/2)}{\tan(k_1 \text{t}/2)}  = -\frac{4k_1 k_2 q^2}{\left( q^2 -k_2^2\right)^2},  \label{E3c} \\
& \frac{A_2}{B_1} = -\imath \frac{2q k_2 \cos(k_2 \text{t}/2)}{\left(q^2-k_2^2\right) \cos (k_1 \text{t}/2)}, \label{E3d} 
\end{align}  
\end{subequations}
for symmetric longitudinal and vertical transverse wave. The anti-symmetric wave becomes \cite{Graff}
\begin{subequations} \label{E4}
\begin{align} 
&u_x =  \left[A_1 \imath q \sin (k_1 z )  + B_2 k_2 \sin (k_2 z) \right] e ^{-\imath(q x - \Omega t)},  \label{E4a} \\
& u_z = \left[ A_1 k_1  \cos (k_1 z )  + B_2 \imath q \cos (k_2 z) \right] e ^{-\imath(q x - \Omega t)}, \label{E4b} \\
& \frac{\tan(k_2 \text{t}/2)}{\tan(k_1 \text{t}/2)}  = -\frac{\left( q^2 -k_2^2\right)^2}{4k_1 k_2 q^2}, \label{E4c} \\
& \frac{A_1}{B_2} = \imath \frac{2q k_2 \sin(k_2 \text{t}/2)}{\left(q^2-k_2^2\right) \sin (k_1 \text{t}/2)}. \label{E4d}
\end{align}  
\end{subequations}
The dispersion relation for longitudinal and vertical transverse wave is then obtained implicitly as 
\begin{align} 
\frac{\tan(k_2 \text{t}/2)}{\tan(k_1 \text{t}/2)}  &= -\left[ \frac{4k_1 k_2 q^2}{\left( q^2 -k_2^2\right)^2} \right]^{\pm 1}, \nonumber \\
 k_1^2 &= \Omega^2/c_1^2 - q^2, \nonumber \\
 k_2^2 &= \Omega^2/c_2^2 - q^2, \label{E5}
\end{align}
where $c_1 = \sqrt{(\lambda+2 \mu)/ \rho}$ is the phase velocity of longitudinal wave. Fig. \ref{F1} shows dispersion curves of both horizontal transverse and longitudinal coupled with vertical transverse waves for an isotropic Silicon plate.

Now, if we carve out a periodic pattern of holes inside this plate, according to Bloch theorem we may expect displacement wave to be of the form
\begin{align} \label{E6}
\mathbf{u}(\mathbf{r}) &= e^{-\imath \bm{q} \cdot \mathbf{r}_{xy}} \tilde{\mathbf{u}}(\mathbf{r}_{xy},z) =  e^{-\imath \bm{q} \cdot \mathbf{r}_{xy}} \sum_{\mathbf{G}} \mathbf{u}_{\mathbf{G}} (z) e^{-\imath \mathbf{G}\cdot \mathbf{r}_{xy} } \nonumber \\
&= \sum_{\mathbf{G}} \left[ u_{\mathbf{G}_{x}} (z) \hat{x} + u_{\mathbf{G}_{y}} (z) \hat{y} + u_{\mathbf{G}_{z}} (z) \hat{z} \right] e^{-\imath (\bm{q} + \mathbf{G}) \cdot \mathbf{r}_{xy} } ,
\end{align} 
where $\tilde{\mathbf{u}}(\mathbf{r}_{xy},z)$ is a periodic function of $x$ and $y$ for every value of $-t/2 < z < t/2$,  $\bm{q}$ is the Bloch wave vector and $\mathbf{G}$ is a reciprocal lattice vector. 

Two types of waves propagate inside a PnC slab: vertical displacement like (VD-like) and horizontal displacement like (HD-like). Displacement of the mid-plane of the slab is perpendicular to slab surface in VD-like and parallel to slab surface in HD-like wave. By adopting wave profile of longitudinal coupled with vertical transverse modes in a plain slab ((\ref{E3a},b) for HD-like and (\ref{E4a},b) for VD-like) we now approximate displacement components in VD-like modes as
\begin{align} \label{E7}
u_{\mathbf{G}_{x}} (z) &\simeq A_{\mathbf{G}_{x}} \sin (k_{1_{\mathbf{G}}} z ) + B_{\mathbf{G}_{x}} \sin (k_{2_{\mathbf{G}}} z ) ,\nonumber \\
u_{\mathbf{G}_{y}} (z) &\simeq  A_{\mathbf{G}_{y}} \sin (k_{1_{\mathbf{G}}} z ) + B_{\mathbf{G}_{y}} \sin (k_{2_{\mathbf{G}}} z ) ,\nonumber \\
u_{\mathbf{G}_{z}} (z) &\simeq A_{\mathbf{G}_{z}} \cos (k_{1_{\mathbf{G}}} z ) + B_{\mathbf{G}_{z}} \cos (k_{2_{\mathbf{G}}} z ) ,
\end{align}
and in HD-like modes as
\begin{align} \label{E8}
u_{\mathbf{G}_{x}} (z) &\simeq A_{\mathbf{G}_{x}} \cos (k_{1_{\mathbf{G}}} z ) + B_{\mathbf{G}_{x}} \cos (k_{2_{\mathbf{G}}} z ) ,\nonumber \\
u_{\mathbf{G}_{y}} (z) &\simeq A_{\mathbf{G}_{y}} \cos (k_{1_{\mathbf{G}}} z ) + B_{\mathbf{G}_{y}} \cos (k_{2_{\mathbf{G}}} z ) ,\nonumber \\
u_{\mathbf{G}_{z}} (z) &\simeq A_{\mathbf{G}_{z}} \sin (k_{1_{\mathbf{G}}} z ) + B_{\mathbf{G}_{z}} \sin (k_{2_{\mathbf{G}}} z ) .
\end{align}
In these equations we set $k_{1_{\mathbf{G}}}^2= \Omega(|\bm{q}+ \mathbf{G}|)^2/c_1^2 - |\bm{q}+ \mathbf{G}|^2$ and $k_{2_{\mathbf{G}}}^2= \Omega(|\bm{q}+ \mathbf{G}|)^2/c_2^2 - |\bm{q}+ \mathbf{G}|^2$, 
where $\Omega(\bm{q})$ of VD-like mode is obtained from the first anti-symmetric dispersion band and that of HD-like mode from the first symmetric dispersion band of a plain slab with the same material and thickness.
Since $\lambda$ and $\mu$ are position dependent, (\ref{E1}) should be written as
\begin{align} \label{E9}
\sum_{j=1}^3 & \left[ \frac{\partial}{\partial x_i}\left(\lambda \frac{\partial u_j}{\partial x_j}\right) + \frac{\partial}{\partial x_j}\left(\mu \frac{\partial u_i}{\partial x_j}\right) + \frac{\partial}{\partial x_j}\left(\mu \frac{\partial u_j}{\partial x_i}\right) \right] \nonumber \\
& = \rho \ddot{u_i}, \quad (i=1,2,3).
\end{align}
In a compact form it can be written as
\begin{equation*} 
\mathbb{L} \ket{\mathbf{u}(\mathbf{r})} = \Omega^2 \mathbb{R} \ket{\mathbf{u}(\mathbf{r})},
\end{equation*} 
which is a generalized eigenvalue problem. We can find an upper bound for its eigenvalues by replacing $\mathbf{u}(\mathbf{r})$ in 
\begin{equation*} 
\Omega^2 = \frac{\bra{\mathbf{u}(\mathbf{r})}\mathbb{L} \ket{\mathbf{u}(\mathbf{r})}}{\bra{\mathbf{u}(\mathbf{r})}\mathbb{R} \ket{\mathbf{u}(\mathbf{r})}} ,
\end{equation*} 
by its approximate value from (\ref{E6})-(\ref{E8}) and searching for optimum values of $A_{\mathbf{G}}$ and $B_{\mathbf{G}}$ coefficients. It can be shown that the optimum value of these coefficients are obtained by solving the following eigenvalue problem, 
\begin{equation*} 
\mathbb{M} 
\begin{pmatrix}
A_{\mathbf{G}_{x}} \\
B_{\mathbf{G}_{x}} \\
A_{\mathbf{G}_{y}} \\
B_{\mathbf{G}_{y}} \\
A_{\mathbf{G}_{z}} \\
B_{\mathbf{G}_{z}} 
\end{pmatrix}_{\mathbf{G}} + \Omega^2 \mathbb{N}
\begin{pmatrix}
A_{\mathbf{G}_{x}} \\
B_{\mathbf{G}_{x}} \\
A_{\mathbf{G}_{y}} \\
B_{\mathbf{G}_{y}} \\
A_{\mathbf{G}_{z}} \\
B_{\mathbf{G}_{z}} 
\end{pmatrix}_{\mathbf{G}} = 0.
\end{equation*}
In this equation $\mathbb{M}$ and $\mathbb{N}$ are obtained by calculating inner products $\bra{\mathbf{u}(\mathbf{r})}\mathbb{L} \ket{\mathbf{u}(\mathbf{r})}$ and $\bra{\mathbf{u}(\mathbf{r})}\mathbb{R} \ket{\mathbf{u}(\mathbf{r})}$, respectively, and extracting $A_{\mathbf{G}}$ and $B_{\mathbf{G}}$ coefficients. 
The VD-like and HD-like band structure of a hexagonal lattice PnC slab calculated utilizing this method, which we call it variational from now on, is compared with the result of 3D PWE method in Fig. \ref{F2}. It can be seen that both methods give similar bands for lower frequencies (below 20th band). 
This is a special crystal which, as is investigated in \cite{Saeed}, has both photonic and phononic band gaps. Hence it can be used as a basis for phoxonic waveguide.  
The advantage of our method over 3D PWE method is its faster convergence. This fact is illustrated in Fig. \ref{F3}.
{In this figure we have calculated the phase velocities of two modes of the crystal at two high symmetry points of its irreducible Brillouin zone, M$^{(5)}$ (fifth mode at M point) and K$^{(3)}$ (third mode at K point) with our variational and 3D PWE methods and plotted them versus their calculation times. We began with three plane waves along $x-$ and $y-$axes for both 3D PWE and our method, that is we set the bound of $\Sigma$ expression of Bloch wave to $N=1$. The number of plane waves along $z-$axis for PWE method is equal to $x-$axis at each point of the figure. We then increased the number of plane waves by 2, $(N=N+1)$, at each successive points of the plot to obtain more accurate results. The right most points of PWE method are related to $13$ plane waves $(N=6)$ along each axis while the right most points of our method are related to $19$ plane waves $(N=9)$ along $x-$ and $y-$ axes.

The variational method converges faster because the field trial function along $z-$ axis in our method needs just two unknown parameters to be determined while in the standard 3D PWE method more Fourier series terms (unknown parameters) are needed to obtain a trial function which its similarity to the exact eigen function is comparable with our method. In other words, our method uses a smarter trial function along $z-$ axis than 3D PWE method that with fewer unknown parameters can give an acceptable trial function. The fewer unknown parameters our problem has, the smaller the final matrix which its eigen value should be calculated would be and the faster it is prepared.
\begin{figure}[ht]
\centering
\includegraphics[width=3.5in] {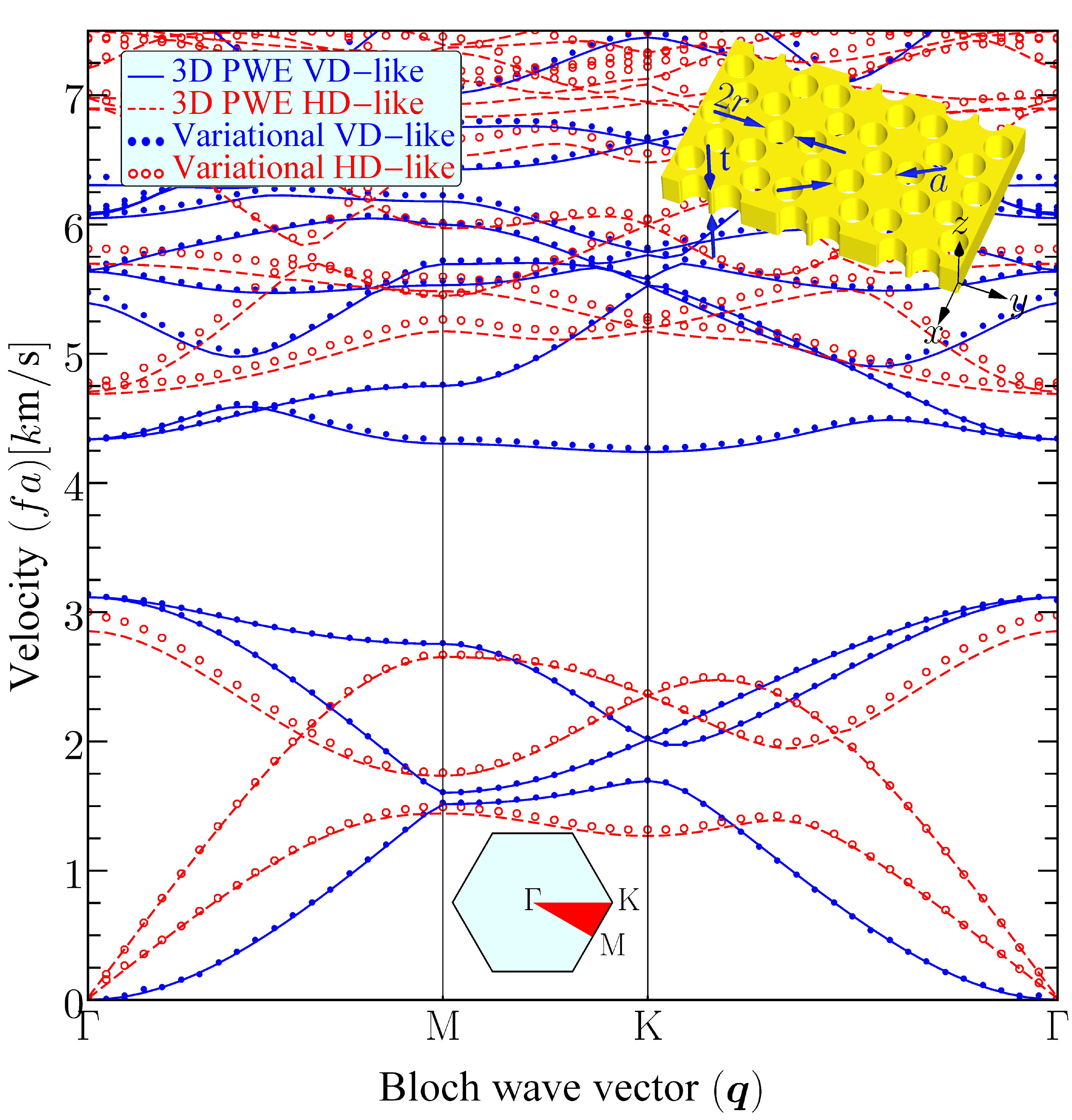}
\caption{ VD-like and HD-like band structures of a slab of PnC with hexagonal lattice. The crystal which can be seen in the upper inset of the figure is made of isotropic Silicon with Lam$\acute{\text{e}}$ constants and mass density $\lambda=44.27G\text{Pa}$, $\mu=80G\text{Pa}$ and $\rho=2329 k\text{g}/\text{m}^{3}$. Slab thickness is $\text{t} = a/\sqrt{3}$ where $a$ is the lattice constant. Radius of vacuum holes equals $r=0.425a/\sqrt{3}$. The lower inset shows the crystal unit cell in reciprocal lattice with the irreducible Brillouin zone highlighted in red. The 3D PWE bands are calculated with $11$ plane waves along each axis, i.e. totally $11^3 =1331$ plane waves. }
\label{F2} 
\end{figure}

\begin{figure}[ht]
\centering
\includegraphics[width=3.5in] {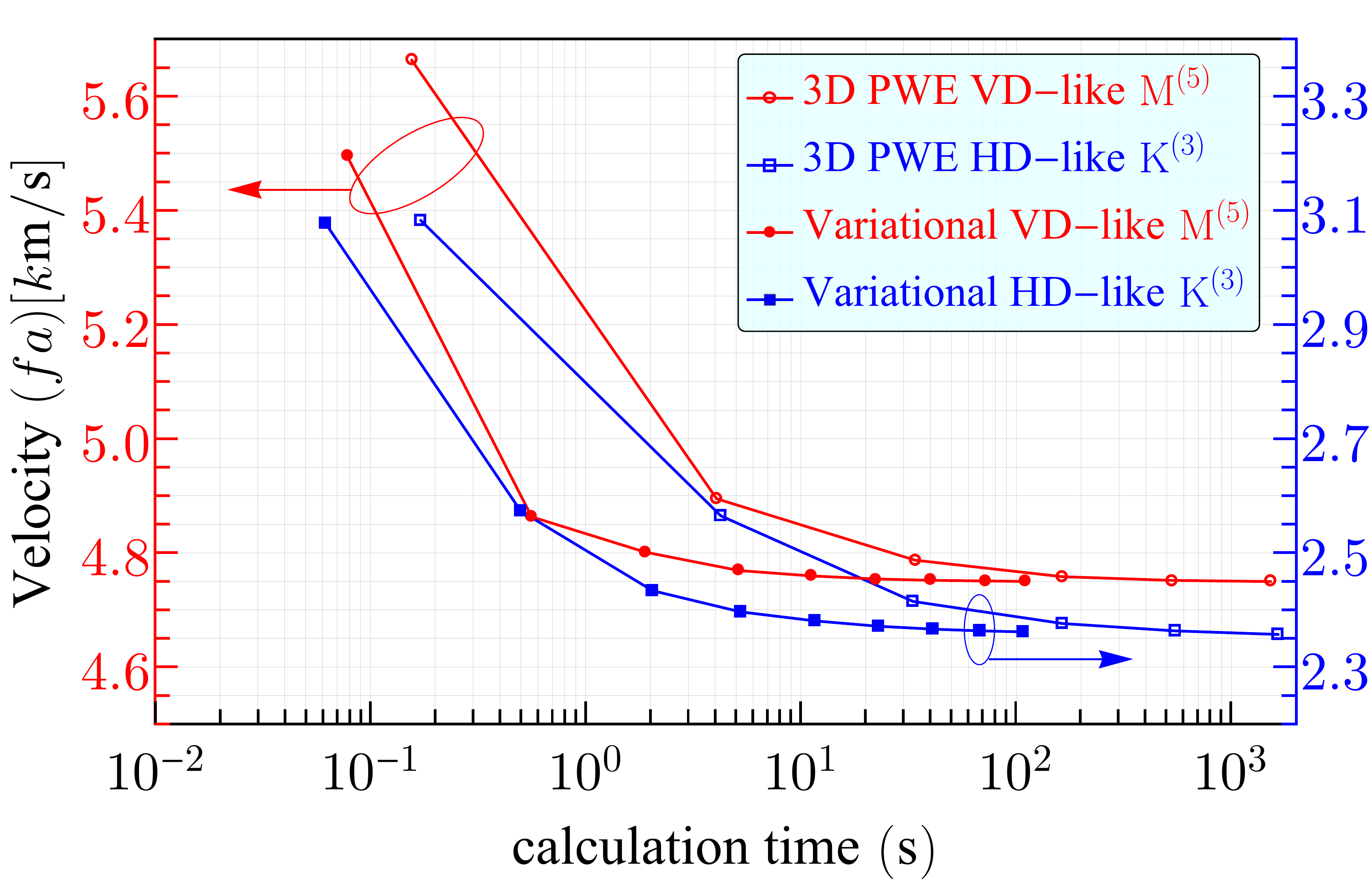}
\caption{3D PWE versus variational convergence speed. The comparison is done in two high symmetry points, $\text{M}^{(5)}$ (fifth mode of  $\text{M}$ point) and $\text{K}^{(3)}$ (third mode of  $\text{K}$ point). The leftmost points are related to three plane waves $(N=1)$ for each direction in plane wave expansion formalism. $N$ has increased by 1 for each point after that. Left and right vertical axes which show the phase velocity of modes are related to red and blue curves respectively. 
All calculations have been performed on a personal computer equipped with an intel i7 processor.}
\label{F3} 
\end{figure}

\subsection{Photonic Crystal Slab}

In a similar manner, we can go through the same process to find a PtC slab modes. Here, the eigenvalue problem is 
\begin{equation} \label{E10} 
\mathbb{K} \ket{\mathbf{H}(\mathbf{r})} = \frac{\omega^2}{c^2}\ket{\mathbf{H}(\mathbf{r})},
\end{equation} 
in which $\mathbf{H}(\mathbf{r})$ is the magnetic field, $c$ is the light velocity in vacuum, $\omega$ is the angular frequency of electromagnetic wave and $\mathbb{K} =\nabla \times \left( \eta(\mathbf{r}) \nabla \times (\cdot) \right)$, where $\eta(\mathbf{r}) = 1/ \epsilon_{\text{r}}(\mathbf{r})$ is the relative impermeability of environment. The magnetic field inside and around a PtC slab can be written as
\begin{align} \label{E11}
\mathbf{H}(\mathbf{r}) &= e^{-\imath \bm{\kappa} \cdot \mathbf{r}_{xy}} \tilde{\mathbf{H}}(\mathbf{r}_{xy},z) =  e^{-\imath \bm{\kappa} \cdot \mathbf{r}_{xy}} \sum_{\mathbf{G}} \mathbf{H}_{\mathbf{G}} (z) e^{-\imath \mathbf{G}\cdot \mathbf{r}_{xy} } \nonumber \\
&= \sum_{\mathbf{G}} \left[ H_{\mathbf{G}_{\|}} (z) \mathbf{e}_{\mathbf{G}_{\|}} + H_{\mathbf{G}_{\perp}} (z) \mathbf{e}_{\mathbf{G}_{\perp}} + H_{\mathbf{G}_{z}} (z) \hat{z} \right]  \nonumber \\
 & \qquad \qquad  \cdot e^{-\imath (\bm{\kappa} + \mathbf{G}) \cdot \mathbf{r}_{xy} } ,
\end{align} 
where we have expanded its elements on the coordinate system of wave vectors. Here, this local coordinate system is represented by the unit vectors lying in the mid-plane of the slab and parallel with (perpendicular to) $\bm{\kappa} + \mathbf{G}$ as $\mathbf{e}_{\mathbf{G}_{\|}}$ ($\mathbf{e}_{\mathbf{G}_{\perp}}$). 
Adopting magnetic field profile around a dielectric slab waveguide, we choose our ansatz magnetic field for the transverse electric like (TE-like) mode as
\begin{align} \label{E12} 
 H_{\mathbf{G}_{\|}} (z) &=  A_{\mathbf{G}_{\|}} f_{\mathbf{G}_{\|}} (z) ,\nonumber \\
 H_{\mathbf{G}_{\perp}} (z) &=  A_{\mathbf{G}_{\perp}} f_{\mathbf{G}_{\perp}} (z),  \nonumber \\ H_{\mathbf{G}_{z}} (z) &= -\imath |\boldsymbol\kappa + \mathbf{G}|  A_{\mathbf{G}_{\|}} f_{\mathbf{G}_{z}} (z),
\end{align} 
where $A_{\mathbf{G}_{\|}}$ and $A_{\mathbf{G}_{\perp}}$ are coefficients to be determined and 
\begin{align}\label{E13}
f_{\mathbf{G}_{\|}}(z) &= \left\{
\begin{array}{lcr}
\sin(k_{\mathbf{G}_{\|}} z)  &  & \qquad \   |z| \le \text{t}/2\\
\sin (k_{\mathbf{G}_{\|}} \text{t}/2) e^{-\alpha_{\mathbf{G}} (z-\text{t}/2)} &  &  z > \text{t}/2 \\
-\sin (k_{\mathbf{G}_{\|}} \text{t}/2) e^{\alpha_{\mathbf{G}} (z+\text{t}/2)} &  &  z < -\text{t}/2
\end{array} \right.  , \nonumber \\ 
f_{\mathbf{G}_{\perp}}(z) &= \left\{
\begin{array}{lcr}
\sin(k_{\mathbf{G}_{\perp}} z)  &  & \qquad \ |z| \le \text{t}/2\\
\sin (k_{\mathbf{G}_{\perp}} \text{t}/2) e^{-\alpha_{\mathbf{G}} (z-\text{t}/2)} &  &  z > \text{t}/2 \\
-\sin (k_{\mathbf{G}_{\perp}} \text{t}/2) e^{\alpha_{\mathbf{G}} (z+\text{t}/2)} &  &  z < -\text{t}/2
\end{array} \right. ,\nonumber \\ 
f_{\mathbf{G}_{z}}(z) &=  \left\{
\begin{array}{lcr}
\cos(k_{\mathbf{G}_{\|}} z) /k_{\mathbf{G}_{\|}}  &  & \quad  |z| \le \text{t}/2\\
\sin (k_{\mathbf{G}_{\|}} \text{t}/2) e^{-\alpha_{\mathbf{G}} (z-\text{t}/2)}/\alpha_{\mathbf{G}} &  &  z > \text{t}/2 \\
\sin (k_{\mathbf{G}_{\|}} \text{t}/2) e^{\alpha_{\mathbf{G}} (z+\text{t}/2)}/\alpha_{\mathbf{G}} &  &  z < -\text{t}/2
\end{array} \right. . \nonumber
\end{align}
In these equations $ \alpha_{\mathbf{G}} = \sqrt{| \boldsymbol\kappa + \mathbf{G}|^2 - \omega(| \boldsymbol\kappa + \mathbf{G}|)^2}$ and $\omega(| \boldsymbol\kappa + \mathbf{G}|)$ is estimated from the first TE dispersion band of a dielectric slab waveguide with the same material and thickness. This estimation is the best one for guided modes, i.e. the modes below the light cone.
$k_{\mathbf{G}_{\|}}$ and $k_{\mathbf{G}_{\perp}}$ are also calculated from the following implicit relations.
\begin{align}
\alpha_{\mathbf{G}}  &=  k_{\mathbf{G}_{\|}} \tan ( k_{\mathbf{G}_{\|}} \text{t}/2), \nonumber \\
\alpha_{\mathbf{G}}  &= -k_{\mathbf{G}_{\perp}} \cot (k_{\mathbf{G}_{\perp}} \text{t}/2) / \epsilon_{r_{\text{eff}}} , \nonumber
\end{align}
where $\epsilon_{r_{\text{eff}}}$ is the effective permittivity of the PtC slab.
These functions are chosen in a manner to satisfy $\nabla \cdot \mathbf{H} (\mathbf{r}) = 0$.
For transverse magnetic like (TM-like) modes, trial functions are similar to the above functions except that
$f_{\mathbf{G}_{\|}}(z)$ and $f_{\mathbf{G}_{\perp}}(z)$ should be even functions and $f_{\mathbf{G}_{z}}(z)$ should be an odd function. Also, $k_{\mathbf{G}_{\|}}$ and $k_{\mathbf{G}_{\perp}}$ are calculated from
\begin{align}
\alpha_{\mathbf{G}}  &=  - k_{\mathbf{G}_{\|}} \cot ( k_{\mathbf{G}_{\|}} \text{t}/2) \nonumber \\
\alpha_{\mathbf{G}}  &= k_{\mathbf{G}_{\perp}} \tan (k_{\mathbf{G}_{\perp}} \text{t}/2) / \epsilon_{r_{\text{eff}}} . \nonumber 
\end{align}
In this mode $\omega(| \boldsymbol\kappa + \mathbf{G}|)$ is estimated from the first TM dispersion band of the dielectric slab waveguide mentioned above.
To test this method, we calculated the band structure of the crystal slab analyzed in the previous sub-section with both 3D PWE and the proposed methods. Fig. \ref{F4a} compares the results.
\begin{figure*}[ht]
\centering
\subfloat[]{\includegraphics[width=3.5in]{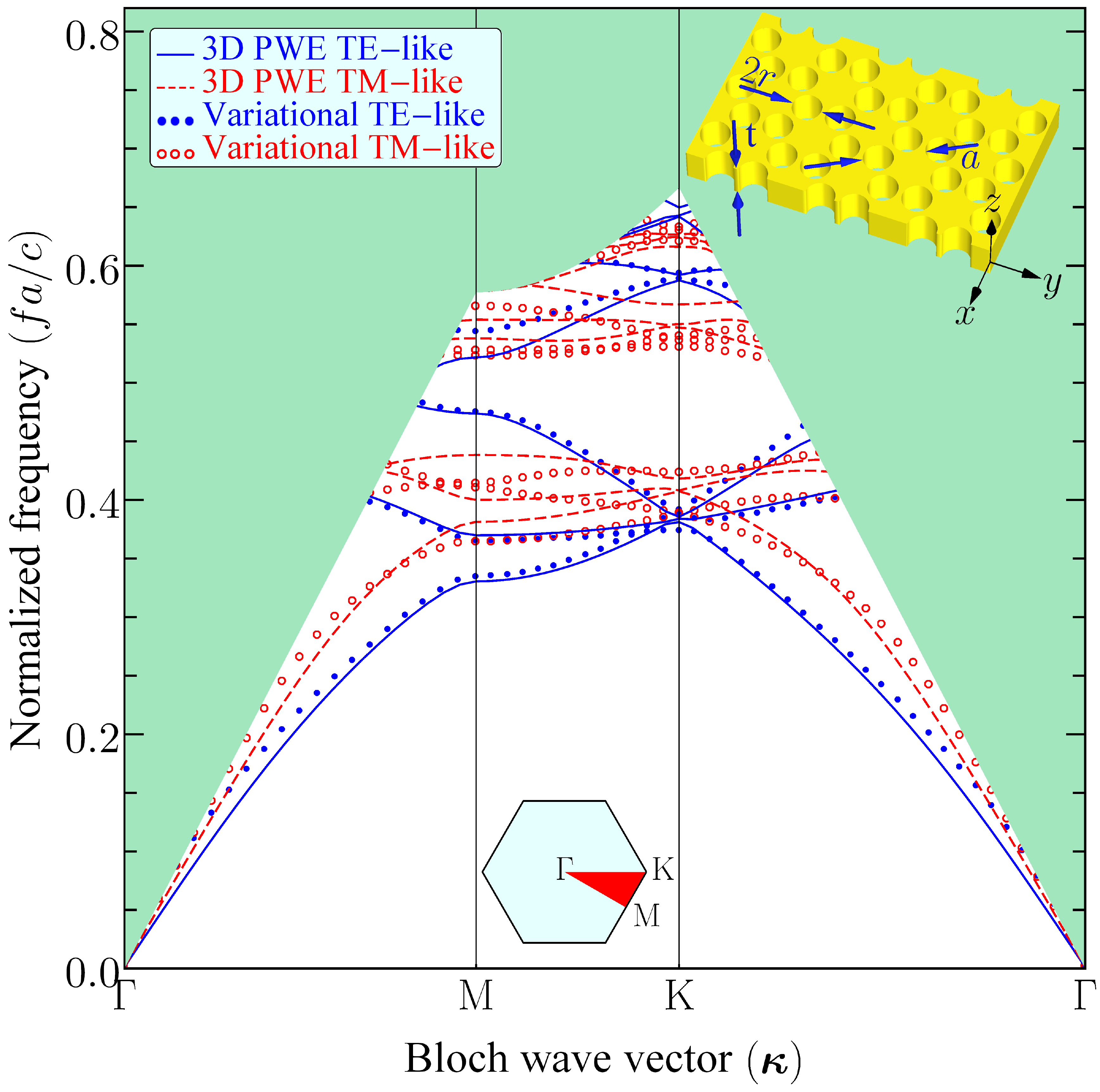}\label{F4a}}
\hspace{2pt}
\subfloat[]{\includegraphics[width=3.5in]{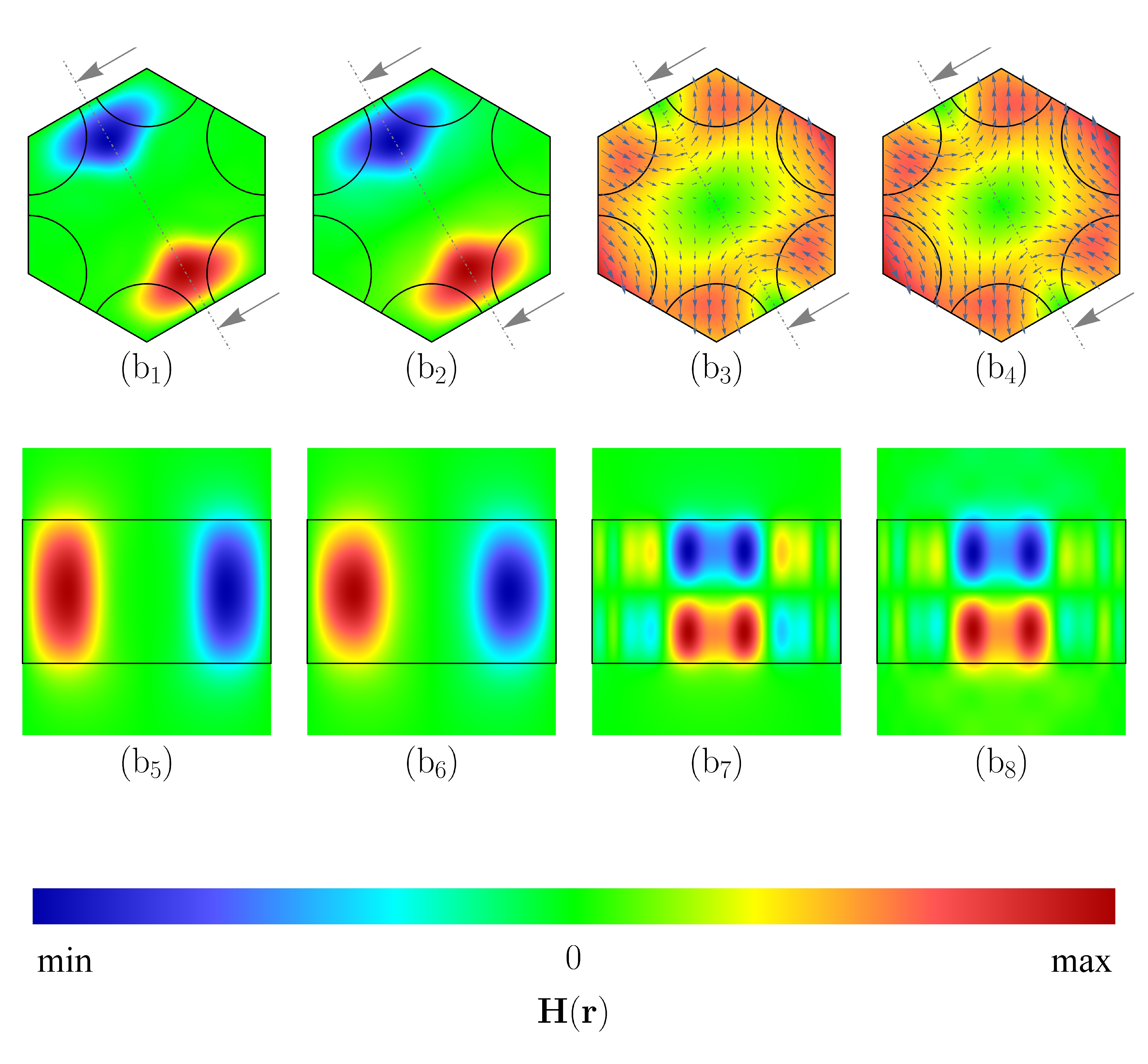}\label{F4b}}
\caption{(a) Band structure of a hexagonal PtC slab made of isotropic Silicon with relative permittivity equals $\epsilon_{\text{r}}=11.9$. The slab thickness is $\text{t} = a/\sqrt{3}$ and the radius of vacuum holes equals $r=0.425a/\sqrt{3}$. Results of both variational and 3D PWE methods are plotted in this figure. Green shaded area shows the light cone and the lower inset shows the crystal unit cell in reciprocal lattice with the irreducible Brillouin zone highlighted in red. The 3D PWE bands are calculated with $11$ plane waves along each axis, i.e. totally $11^3 =1331$ plain waves.
(b) Magnetic field profiles at two high symmetry points of Brillouin zone computed with the variational and 3D PWE methods. Figures $(\text{b}_1)$, $(\text{b}_3)$, $(\text{b}_5)$ and $(\text{b}_7)$ are calculated with variational method and the others are results of PWE. $(\text{b}_1)$ and $(\text{b}_2)$ show TE-like $H_z$ of M$^{(4)}$ (fourth mode in M point) in the mid-plane of the crystal unit cell. $(\text{b}_5)$ and $(\text{b}_6)$  illustrate $H_z$ in the plane perpendicular the to crystal slab surface at the specified cross sections of $(\text{b}_1)$ and $(\text{b}_2)$ respectively. $(\text{b}_3)$ and $(\text{b}_4)$ show TM-like $\mathbf{H}$ of K$^{(3)}$ (third mode in K point) in the mid-plane of the crystal unit cell. $(\text{b}_7)$ and $(\text{b}_8)$ are $H_z$ in the plane perpendicular to the crystal slab surface at the specified cross sections of $(\text{b}_3)$ and $(\text{b}_4)$ respectively.}
\label{F4} 
\end{figure*}

Since both of the methods are based on variational principle, each of them gives a lower frequency for a mode is more accurate in that mode. Inspecting Fig. \ref{F4a}, reveals that our proposed method is more accurate in plenty of modes. In TE-like mode, our method results in bands which are very similar in shape to that of 3D PWE, while in TM-like mode, shapes of bands differ slightly. However, our method is more accurate in TM-like than TE-like modes. 

In Fig. \ref{F4b} we have compared the mode profiles calculated with variational and 3D PWE methods at two high symmetry points of Brillouin zone. It can be seen the results are very similar.

%%%%%%%%%%%%%%%%%%%%%%%%%%%%%
\section{Photo-elastic Interaction in Slab Optomechanical Waveguides}

%To illustrate an application of the proposed method we first need to explain how elastic and electromagnetic waves interact inside an optomechanical waveguide . 

Interaction of electromagnetic and elastic waves originates from different mechanisms among them are electrostriction, magnetostriction, radiation pressure, piezoelectricity, photoelasticity, and interface displacement \cite{Pennec3}. Electrostriction, magnetostriction and radiation pressure are mechanisms through which electromagnetic wave affects elastic wave. In piezoelectric materials electromagnetic and elastic waves can have mutual effects on each other through piezoelectricity phenomenon. The last two mechanisms, i.e. photoelasticity and interface displacement are responsible for effects of elastic wave on electromagnetic wave \cite{Eichenfield2009, DjafariRouhani2014}. In general, for piezo-electric nano-structured materials, radiation pressure terms which arise from moving boundaries \cite{Djafari,Balram,Khorasani} are negligible compared to the bulk electrostrictive forces arising from photo-elastic interactions. For non-piezoelectric media, radiation pressure and electrostriction have been shown to contribute comparably, or at least within the same order of magnitude \cite{Djafari}. A coupling coefficient between electromagnetic and elastic waves can be defined for each of these mechanisms. Different expressions have been presented for the coupling coefficient originated from photoelastic effect} \cite{Eichenfield2009, DjafariRouhani2014, Khorasani}. In this section we present a new expression for this coefficient which can give some insight about electromagnetic wave evolution in time in the presence of elastic wave. We use this expression to illustrate an application of our proposed method.

Now suppose we have an optomechanical slab waveguide in which both elastic and electromagnetic wave propagate along $x-$axis. We start with (\ref{E10}) but with a perturbed operator as
\begin{equation} \label{E13}
(\mathbb{K} + \Delta \mathbb{K}) \mathbf{H}'(\mathbf{r}, t) = -\frac{1}{c^2} \frac{\partial ^2}{\partial t^2}\mathbf{H}'(\mathbf{r}, t),
\end{equation} 
in which $\Delta \mathbb{K}= \nabla \times \left( \Delta \bm{\eta}(\mathbf{r}, t) \nabla \times (\cdot) \right)$ 
and 
\begin{equation} \label{E14}
\Delta \bm{\eta} = \bm{P}: \bm{\varepsilon} \Longleftrightarrow \Delta \eta_{ij} = \sum_{k,l = 1}^3 P_{ijkl} \varepsilon_{kl}.
\end{equation}
In (\ref{E14}), $\Delta \bm{\eta}$ is the change in impermeability tensor caused by strain wave $ \bm{\varepsilon}= \left[ (\nabla \mathbf{u})^T + \nabla \mathbf{u} \right]/2 $ and $ \bm{P}$ is the photo-elastic tensor. With regards to (\ref{E6}) we can write $\Delta \bm{\eta}$ as 
\begin{equation} \label{E15}
\Delta \bm{\eta} (\mathbf{r}, t) = \bm{v} (\mathbf{r}) e^{-\imath (q x - \Omega t)},
\end{equation}
where $\bm{v} (\mathbf{r}) $ is a periodic tensor of $x$ and $q$ is the elastic wave number. Since $\mathbb{K}$ is a Hermitian operator, its eigenstates build a complete set on which we can expand $\mathbf{H}'(\mathbf{r}, t)$ as
\begin{align} \label{E16}
\mathbf{H}'(\mathbf{r}, t) =  \sum_{\kappa,n} \alpha_{n, \kappa} (t) \tilde{\mathbf{H}}_{n, \kappa} (\mathbf{r}_{xy},z) e^{-\imath (\kappa x - \omega_{n, \kappa} t)},
\end{align}
where the sum is over all wave numbers in the irreducible Brillouin zone and all eigenstates corresponding to each value of wave number. Here, we have discretized the wavenumber just for the sake of simplicity. It can be replaced by an integral as 
$ \left( \sum_{\kappa} \longleftrightarrow \frac{a}{\pi} \int_{0}^{\pi/a} \ud \kappa \right) $, where $a$ is the length of our crystal slab waveguide unit cell along $x$ axis.
$\alpha_{n, \kappa} (t)$ are time-dependent constants to be determined. We now insert $\mathbf{H}'(\mathbf{r}, t)$ from (\ref{E16}) into (\ref{E13}) to obtain
\begin{align} \label{E17}
& \Delta \mathbb{K}  \sum_{\kappa,n} \alpha_{n, \kappa} (t) \tilde{\mathbf{H}}_{n, \kappa} (\mathbf{r}_{xy},z) e^{-\imath (\kappa x - \omega_{n, \kappa} t)}   \nonumber \\
& = -\frac{1}{c^2} \sum_{\kappa,n}  \left[ 2 \imath \omega_{n, \kappa} \dot{\alpha}_{n, \kappa} (t) + \ddot{\alpha}_{n, \kappa} (t) \right]   \tilde{\mathbf{H}}_{n, \kappa} (\mathbf{r}_{xy},z) \nonumber \\
& \qquad \qquad \qquad \qquad \qquad \qquad \qquad \cdot e^{-\imath (\kappa x - \omega_{n, \kappa} t)} \nonumber \\
& \simeq -\frac{1}{c^2}  \sum_{\kappa,n}  2 \imath \omega_{n, \kappa} \dot{\alpha}_{n, \kappa} (t)   \tilde{\mathbf{H}}_{n, \kappa} (\mathbf{r}_{xy},z)  e^{-\imath (\kappa x - \omega_{n, \kappa} t)} . 
\end{align} 
The second equation is obtained with the assumption, $ \dot{\alpha}_{n, \kappa} (t) \ll \omega_{n, \kappa}$. 
By inner producting both sides of (\ref{E17}) by a specific eigenstate of $\mathbb{K}$ such as $ \tilde{\mathbf{H}}_{m, \ell} (\mathbf{r}_{xy},z)  e^{-\imath (\ell x - \omega_{m, \ell} t)} $ we get
\begin{align} \label{E18}
& \frac{1}{l_x} \int\limits_{\text{V}} \sum_{\kappa,n} \alpha_{n, \kappa} (t) \tilde{\mathbf{H}}^*_{m, \ell} (\mathbf{r})  e^{\imath (\ell x - \omega_{m, \ell} t)} \nonumber \\
& \qquad \qquad \qquad \cdot \Delta \mathbb{K} \left[ \tilde{\mathbf{H}}_{n, \kappa} (\mathbf{r}) e^{-\imath (\kappa x - \omega_{n, \kappa} t)} \right] \ud \mathbf{r} \nonumber \\
& \simeq -\frac{1}{c^2}  2 \imath \omega_{m, \ell} \dot{\alpha}_{m, \ell} (t)  .
\end{align} 
In this equation, $V$ is the volume in which we take the integral and should be enlarged to occupy the whole space and $l_x$ is the dimension of $V$ along the $x-$axis. We also have supposed the eigenstates to be normalized, i.e. 
\begin{equation*} 
\frac{1}{l_x} \int\limits_{\text{V}} \tilde{\mathbf{H}}^*_{m, \ell} (\mathbf{r})  e^{\imath \ell x} \tilde{\mathbf{H}}_{n, \kappa} (\mathbf{r}) e^{-\imath \kappa x } \ud \mathbf{r} = \delta_{n,m} \delta_{\kappa, \ell} ,
\end{equation*}
where $\delta_{n,m}$ and $ \delta_{\kappa, \ell}$ are Kronecker delta functions. By affecting $\Delta \mathbb{K}$, the equation becomes
\begin{align} \label{E19}
 &- \frac{2\imath}{c^2} \omega_{m, \ell} \dot{\alpha}_{m, \ell} (t)  \nonumber \\ 
 & \simeq \frac{1}{l_x} \int\limits_{\text{V}} \sum_{\kappa,n} \alpha_{n, \kappa} (t) \tilde{\mathbf{H}}^*_{m, \ell} (\mathbf{r}) e^{ \imath \ell x} \cdot \nabla \times \Big[\bm{v}(\mathbf{r}) e^{ -\imath q x} \nabla \times  \nonumber \\
& \qquad \qquad \qquad \left( \tilde{\mathbf{H}}_{n,\kappa} (\mathbf{r}) e^{ -\imath \kappa x} \right) \Big] e^{ \imath (\omega_{n, \kappa}-\omega_{m, \ell} + \Omega) t } \ud \mathbf{r} \nonumber \\
&= \sum_{\kappa,n} \xi_{m,n,\ell, \kappa} \alpha_{n, \kappa} (t) e^{ \imath (\omega_{n, \kappa}-\omega_{m, \ell} + \Omega) t } ,
\end{align} 
where 
\begin{align} 
&\xi_{m,n,\ell, \kappa} = \frac{1}{l_x} \int\limits_{\text{V}} \tilde{\mathbf{H}}^*_{m, \ell} (\mathbf{r}) e^{ \imath \ell x} \cdot \nonumber  \\
& \qquad \qquad \quad  \nabla \times \Big[\bm{v}(\mathbf{r}) e^{ -\imath q x} \nabla \times \left( \tilde{\mathbf{H}}_{n,\kappa} (\mathbf{r}) e^{ -\imath \kappa x} \right) \Big]   \ud \mathbf{r}, \nonumber
\end{align}
is the coupling strength between $(\ell, m)$ and $(\kappa, n)$ modes through photoelastic effect. Replacing $\alpha_{m, \ell} (t)$ by $ \beta_{m, \ell} (t) =  \alpha_{m, \ell} (t) e^{ \imath \omega_{m, \ell}  t}$ and doing some simplifications, (\ref{E19}) becomes 
\begin{align} \label{E20}
\dot{\beta}_{m, \ell} (t) &=  \imath \frac{c^2}{2\omega_{m, \ell} } \sum_{\kappa,n} \xi_{m,n,\ell, \kappa} \beta_{n, \kappa} (t) e^{ \imath \Omega t } \nonumber \\ 
&  + \imath \omega_{m, \ell} \beta_{m, \ell} (t) , 
\end{align} 

(\ref{E20}) is a system of ordinary differential equations. Suppose the system is initially in $(\kappa_1, 1)$ state, then the stronger the coupling between an arbitrary state and this initial state, the faster the growth of its coefficient in time. The first step in solving this system of equations is to calculate the coupling strength between different states. In the next subsection, we utilize our proposed method in previous sections for this purpose to demonstrate its usefulness in computing the optomechanical coupling of photonic and phononic modes \cite{Khorasani}.

%%%%%%%%%%%%%%%%%%%%%%%%%%%%%
\subsection{Application Example}

\begin{figure*}[!t]
\centering
\subfloat[]{\includegraphics[width=2in]{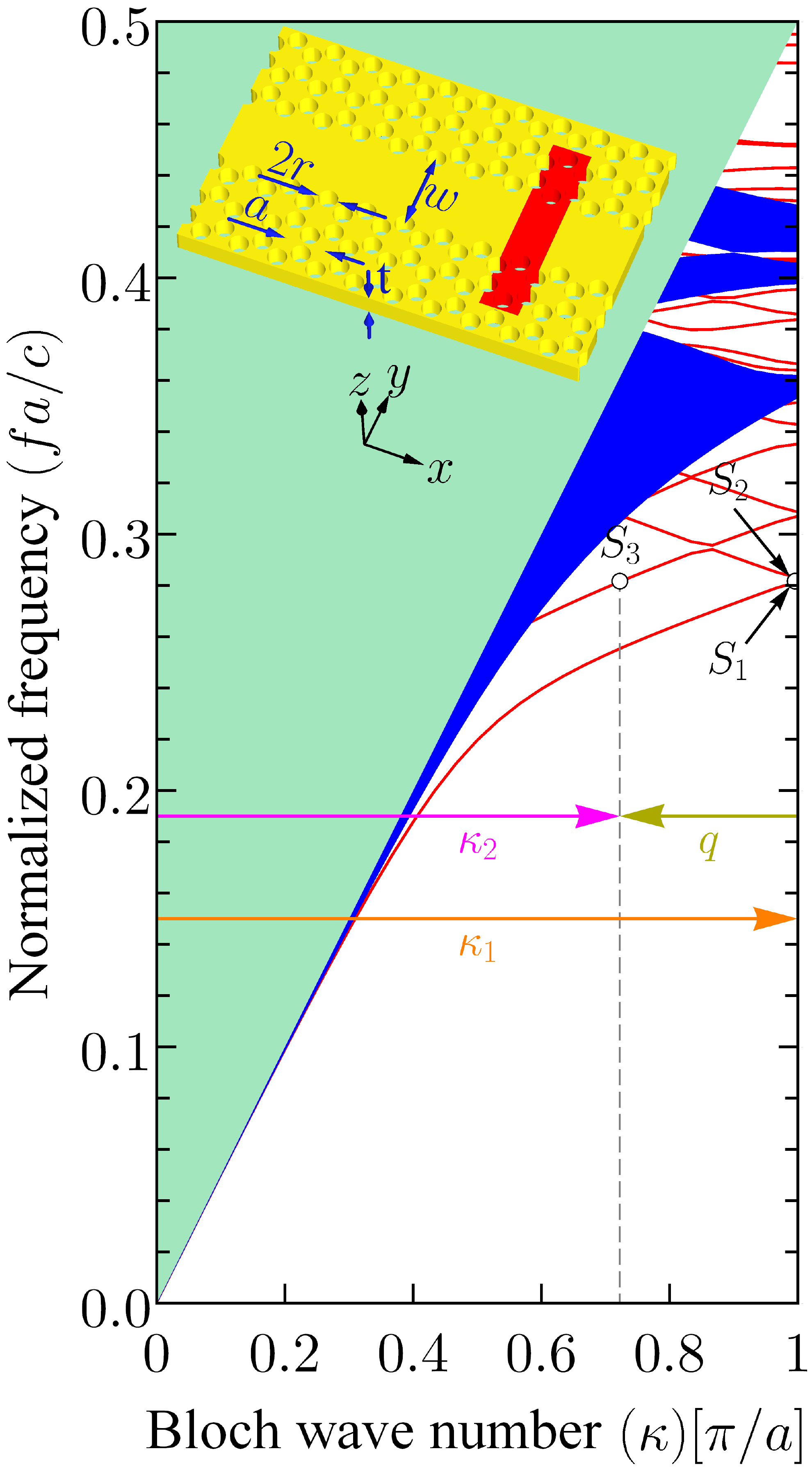}\label{F5a}}
\hspace{2pt}
\subfloat[]{\includegraphics[width=2in]{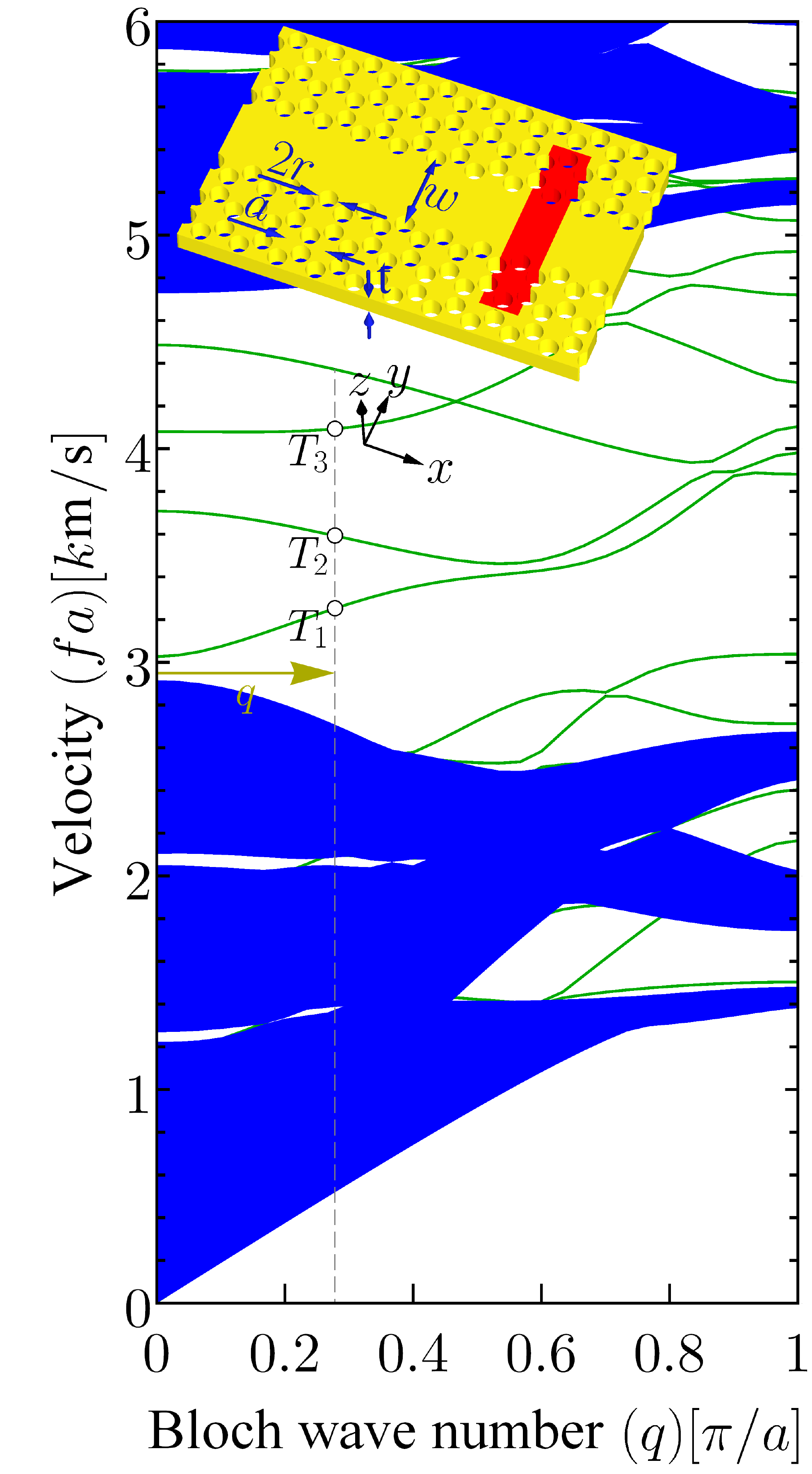}\label{F5b}}
\hspace{16pt}
\subfloat[]{\includegraphics[height=3.6in]{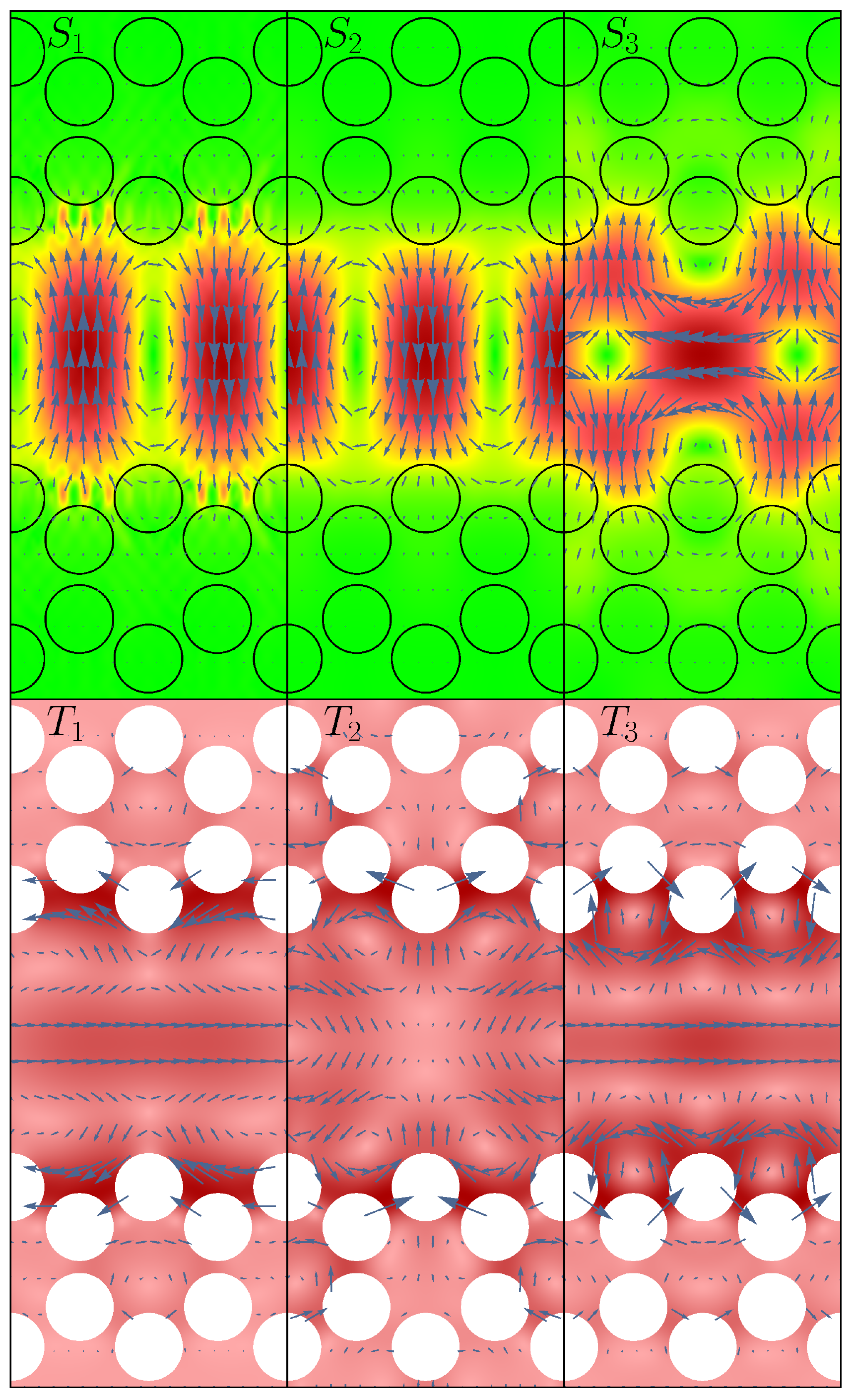}\label{F5c}}
\hspace{1pt}
\subfloat{\includegraphics[width=0.5in]{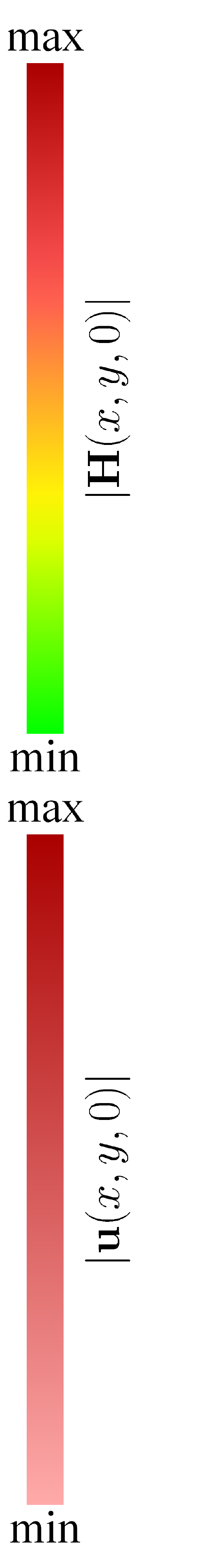}} 
\caption{Band structure of the intended phoxonic waveguide for (a) photonic TM-like and (b) phononic HD-like modes. The waveguide shown in the inset of (a) and (b) is created by introducing a gap with width $w=(1.5+ 1/\sqrt{3}) a$ along $\Gamma \text{K}$ direction in the hexagonal crystal studied in previous sections. The red region in the waveguide of inset of (a) and (b) illustrate the unit cell considered for the waveguide analysis. Green shaded area in (a) shows the light cone and the blue regions are the extended modes of the crystal. Three photonic and three phononic modes are marked in (a) and (b) respectively for coupling calculations. The wave profile of these modes in the mid-plane ($z=0$) of the crystal slab is plotted in (c). The difference in wave number of $S_1$ and $S_3$ equals the wave number of $T_1$, $T_2$ and $T_3$.}
\label{F5}
\end{figure*}

To demonstrate one of the applications of our modified PWE method, we compute the coupling strength due to photoelasticity between different modes of a PtC slab waveguide caused by phononic modes. %The coupling is due to the elastic scattering of optical waves by elastic waves, which cause frequency up- and down-conversion with scattered wavevectors. 
This phonomenon is at the heart of optomechanical interactions in phoxonic crystals  and can be theoretically investigated by coupled-mode theory once the propagating eigenmodes of the slab are known \cite{Khorasani}. Conventional PWE method is inappropriate here since it is too much time and memory consuming, and also suffers from non-uniform convergence. It means that the numerical solutions start to oscillate around the values ideally obtained at infinite number of expansion terms instead of exponential convergence \cite{Aram}. At first, we need to design a simultaneous photonic and phononic (phoxonic) waveguide. For this purpose we use the hexagonal crystal slab analyzed in previous sections. As can be seen in Figs. \ref{F2} and \ref{F4}, this crystal has both photonic and phononic forbidden gaps in its band structures, so it is a good candidate to be used as the base of our waveguide. By inserting a gap in the mentioned crystal along $\Gamma \text{K}$ direction, and following the design instructions given elsewhere \cite{Khorasani}, we obtain a phoxonic waveguide as is shown in the inset of Fig. \ref{F5a}.
 
It can be proved that VD-like elastic waves cannot couple any two optical modes in a phoxonic slab waveguide through photoelasticity, i.e. the coupling strength between any two optical modes through a VD-like elastic wave equals zero. This is actually quite expected from overlap integrals of optomechanical interactions \cite{Khorasani}. 
Hence, we calculated photonic TM-like and phononic HD-like band structures of the designed waveguide using our proposed variational method in this work. Figs. \ref{F5a} and \ref{F5b} show these band structures.
To compute the optomechanical coupling strength we need to obtain $\Delta \bm{\eta}$. We used the photo-elastic tensor of Silicon reported in \cite{Biegelsen} for this purpose. The independent photo-elastic tensor elements of Silicon are $p_{1111}=-0.094$, $p_{1122}=0.017$ and $p_{1212}=-0.051$. The coupling strengths $\xi_{S_1, S_3}$ and $\xi_{S_2, S_3}$ through phononic modes $T_1$, $T_2$ and $T_3$ are gathered in Table \ref{T1}. These values are calculated after normalizing phononic waves profiles, i.e.
\begin{equation*} 
\frac{1}{a^3}\int_{-\text{t}/2}^{\text{t}/2} \int_{-\infty}^{\infty} \int_{0}^a |\mathbf{u}(\mathbf{r})|^2 \ud x \ud y \ud z = a^2 .
\end{equation*}
The specified modes are marked in Figs. \ref{F5a} and \ref{F5b}.  

\begin{table}[!ht]
\renewcommand{\arraystretch}{1.3}
\caption{Coupling Strength Between Optical Waves Through Elastic Waves due to Photoelastic Effect}
\label{T1}
\centering
\begin{tabular}{c c c}
\noalign{\hrule height 1pt}
 & $\xi_{S_1, S_3}$ & $\xi_{S_2, S_3}$ \\
\hline
$T_1$ & $0.9287 + \imath 0.0875$  & $0$  \\
\hline
$T_2$ & $-0.0046 + \imath 0.7713$ & $0$  \\
\hline
$T_3$ & $0$ & $0.1011 - \imath 0.2290$  \\
\noalign{\hrule height 1pt}
\end{tabular}
\end{table}

\section{Conclusion}
A new approach for analysis of photonic and phononic crystal slab was proposed. This approach which is based on PWE method converges much faster than the conventional 3D PWE method and is uniformly convergent. Furthermore, it lets us analyze crystal slabs with large unit cells by a personal computer unlike 3D PWE method which needs a huge amount of memory in such cases. An application example was presented, where we constructed a non-reciprocal photonic device utilizing a phoxonic waveguide and calculated the coupling strength between different modes with the proposed method. Further details and deep discussions on this subject are being intended as a separate article to be published soon \cite{Aram2017}.

\ifCLASSOPTIONcaptionsoff
  \newpage
\fi



\bibliographystyle{IEEEtran}
\bibliography{Ref}

%\begin{IEEEbiographynophoto}{Mohammad Hasan Aram} was born in Tehran, Iran, in 1987. He received his B.Sc., M.Sc. and Ph.D. degrees all in Electrical Engineering from Sharif University of Technology, Tehran, in 2009, 2011 and 2015, respectively. He is currently a Post-doctoral researcher with the School of Electrical Engineering at Sharif University of Technology. His fields of interest include photonics, quantum optics, quantum information, quantum computing and opto-mechanics.

%\begin{IEEEbiographynophoto}{Sina Khorasani (S'98-M'05-SM'09)} received his M.Sc. and Ph.D. degrees from Sharif University of Technology respectively in 1996 and 2001, both in Electrical Engineering, where he is a Full Professor. He has been with School of Electrical and Computer Engineering at Georgia Institute of Technology as a Postdoctoral (2002-2004) and Research Fellow (2011-2012). He is now with \'{E}cole Polytechnique F\'{e}d\'{e}rale de Lausanne as a Visiting Professor. His active research areas include quantum optics and photonics, and quantum electronics. Dr. Khorasani is a Senior Member of IEEE.

\end{document}